\documentclass[prb,aps,showpacs,amsmath]{revtex4}

\newcounter{mycount}

\newcommand{\be}[1]{ \begin{eqnarray} \mbox{$\label{#1}$} }
\newcommand{\ee}{\end{eqnarray}}
\newcommand{\pref}[1]{(\ref{#1})}
\newcommand\ie {{\it i.e. }}
\newcommand\eg {{\it e.g. }}

\newcommand\half{\frac 1 2 }

\newcommand{\bracket}[2]   {  \left<#1 |  #2\right>}
\newcommand{\av}[1]{\langle #1\rangle}

\newcommand\noi{\noindent}
\newcommand{\p}{\partial}

\newcommand{\rmi}{{\mathrm i}}
\newcommand{\rmd}{{\mathrm d}}
\newcommand{\rme}{{\mathrm e}}
\newcommand{\zb}{\bar z}

\newcommand\vphi{\varphi}
\newcommand\vphib{\bar\varphi}

\newcommand{\xib}{\bar\xi}
\def\a{\alpha}
\def\b{\beta}

\begin{document}

\title{Quasihole condensates in quantum Hall  liquids}

\author{J. Suorsa$^1$}
\author{S. Viefers$^1$}
\author{T.H. Hansson$^2$}

\affiliation{
$^1$Department of Physics, University of Oslo, Box 1048 Blindern, NO-0316 Oslo, Norway \\
$^2$Department of Physics, Stockholm University, AlbaNova University Center, SE-106 91 Stockholm, Sweden }

\date{\today}

\begin{abstract} 
We develop a formalism to describe quasihole condensates in quantum Hall  liquids and thereby extend the conformal field theory approach to  the full hierarchy of spin-polarized Abelian states, and to several classes of non-Abelian hierarchical states. Most  previously proposed spin-polarized quantum Hall wave functions appear as special cases. In this paper we explain the physical motivations for the approach, and exemplify it by explicitly constructing the level-two  quasihole condensate state at filling fraction 2/3, and the two level-three states at 5/13 and 5/7 which are built from combinations of quasielectron and quasihole condensates.
\vskip 3mm \noi

\end{abstract}


\maketitle

\section{Introduction}

Ever since the fractional quantum Hall effect (FQHE) was discovered in 1982\cite{fqhe}, the quantum Hall system has been subject of intense interest from both experimentalists and theorists. From a theoretical point of view, the FQH states fall into two main groups, the Abelian and the non-Abelian, depending on the type of statistics obeyed by their quasiparticles. All the states observed in the lowest Landau level (LLL) can naturally be classified as Abelian, while there are second Landau level states which are strong candidates for being non-Abelian. The present surge in interest has been focused on the non-Abelian states, because of their tentative importance for quantum information technology, but we are also still far from having a comprehensive description of the Abelian states. 

To understand the plethora of observed FQH states, theorists have used a number of different approaches. A very successful one has been the construction of explicit wave functions -- a craft that was initiated by Laughlin\cite{laugh}, and has been pursued ever since. Prominent examples are the hierarchical wave functions proposed by Haldane\cite{haldhi} and Halperin\cite{halphi}, the composite fermion wave functions constructed by Jain\cite{jain}, and wave functions derived from correlators in conformal field theory (CFT)\cite{mr}, which are the subject of this paper. 

An explicit wave function does not necessarily add much to our understanding if it does not fit into a more comprehensive theoretical scheme that can be, if not derived, at least motivated by microscopic physics. Unfortunately several of  the proposed schemes for constructing wave functions are not doing too well on this account. They are either based on strong phenomenological assumptions, such as formation of composite fermions, or provide a general framework, rather than explicit predictions, as in the case of the hierarchical constructions of quasiparticle condensates. It is significant that the intermediate between a fully microscopic description in terms of wave functions, and the extreme low-energy theory in terms of topological Chern-Simons theory, is to a great extent missing. Only in the simplest cases such as the Laughlin states at $\nu = 1/m$ and the Halperin bilayer states\cite{halphi} are there proposals for effective dynamical low-energy theories of the Ginzburg-Landau Chern-Simons (GLCS) type that can be microscopically motivated.\footnote{
Such a theory has also been proposed for the  bosonic Pfaffian state in Ref. \onlinecite{bosepfaff}, but in this case there is no microscopic derivation. The theory has rather been constructed as to have the correct topological field theory as its extreme low-energy limit.}
In both these cases, one recovers a topological Chern-Simons theory in the extreme low-energy limit,  while at the same time one can derive the actual electronic  wave functions. In this connection we should also mention that there exists a field theoretical description of composite fermions\cite{fradkin}.

The CFT based approach pursued in this paper is at an intermediate level between a dynamical effective field theory and a fully microscopic picture. As is briefly described in the next section, it is closely related to the low-energy Chern-Simons theories, and just as the GLCS theories, it provides concrete representative wave functions. However, it differs from the GLCS theories in that it does not specify any dynamics for the bulk modes.  Rather, it is topological in the sense that it correctly describes charges and statistics of the bulk quasiparticles as well as the number and character (charge, chirality) of the edge modes. Furthermore, it captures the response of the quasiparticles to the local curvature of the surface on which the QH liquid lives, via Berry phases related to the orbital spin of the constituent particles. While this response is geometrical rather than topological, it is closely connected to a topological quantity known as the shift, whose value on the sphere is directly related to  a transport coefficient, the QH viscosity\cite{read08}. 
 
We have developed methods that allow us to explicitly construct representative wave functions for arbitrary states in the spin-polarized Abelian hierarchy, and also for several classes of hierarchical non-Abelian states. These include all states based on composite fermions as well as the proposed hierarchy based on the Moore-Read state in the second Landau level\cite{BS}. In this paper we discuss the physical motivations for our proposal and present several explicit examples. In a forthcoming paper\cite{coming} we will present the general theory and provide technical details that are omitted here.

An important ingredient of the CFT framework is that the quasielectrons are not described by local operators, but quasi-local ones, smeared over the distance of a magnetic length\cite{hhv08,hhv09}. The physical reason, which is well is understood, will be reviewed below. In this paper we show that in order to describe hierarchical states that contain condensates of quasiholes, both the electrons and quasiholes are most naturally represented by quasi-local operators. In particular, we develop a two-fluid picture of a general hierarchical state. One of the components consists of the original electrons, and all the quasielectron condensates, while the other consists of the quasihole condensates. The first component is described by chiral and the second by anti-chiral CFT operators. These components can be thought of as two oppositely charged QH liquids  with strong intercorrelations. The latter are manifested in that an electron at position $z$ has both a positive and a negative charge component smeared over a distance of a magnetic length around $z$. The formalism underlying these words will be developed in this paper. Alternatively, the state can be described as a strongly intercorrelated state of a single electrically charged Laughlin-type liquid and a purely topological liquid that is physically manifested in the neutral edge modes. The wave functions obtained in these two descriptions differ, but we believe they describe the same topological phases of matter. 

Sections \ref{sec:bg} and \ref{cfthier} contain background material needed to explain notation and make the paper reasonably self contained. Readers familiar with our earlier work in Refs. \onlinecite{hansson07}, \onlinecite{hhv08} and \onlinecite{hhv09}, can proceed directly to section \ref{sec:antihol} which discusses the use of anti-chiral CFT operators, and anti-holomorphic conformal blocks, in the simplest case of the Laughlin states. Section \ref{twocomp}, which is the most central in the paper, explains how to give a CFT description of hierarchical states that involve quasihole condensates. Although the emphasis in this paper is on the Abelian hierarchy states, the methods can also be applied to non-Abelian states, and in  section \ref{sec:NA} we briefly discuss some preliminary work on this subject. Section \ref{sec:sum} contains a short summary of the main results and a discussion of open questions.


\section{Conformal Field Theory and the Quantum Hall Effect}
\label{sec:bg}

\noi
In order to put our work into context, and make the paper reasonably self-contained, we start out by summarizing in a non-technical way some of the central aspects of the CFT-QHE connection. We shall concentrate on the Abelian hierarchy of fully polarized quantum Hall states, which is the main topic of this paper.


\subsection{General considerations}

\noi
Although there is no rigorously proven connection between quantum Hall wave functions and correlators in conformal field theories (CFTs), there are suggestive heuristic arguments, and at a practical level many successful QH wave functions can indeed be expressed in the language of CFT. The main arguments underpinning the original suggestions of a QH-CFT connection given by Moore and Read\cite{mr}, and by Wen\cite{wencft}, are as follows. First, there are strong general reasons to believe that the low-energy description of a 2D electron gas in a strong magnetic field should be a Chern-Simons theory. In the simplest case of the Laughlin fluids, this can be derived from a microscopic description using mean field assumptions\cite{cstheory,fradkin}, but there are also some quite general arguments for this proposition\cite{froh}. 

There are two different, but closely related, ways to connect the Chern-Simons theory to a CFT. The first is based on results by Witten, who showed that the expectation values of braided and knotted Wilson lines in a Chern-Simons theory are directly manifested as monodromies of certain conformal blocks\cite{witt}, which are related to correlators of conformal fields in a corresponding CFT. Physically these monodromies describe the statistical phases resulting from braiding of the anyonic quasiparticles in the QH fluids. Since the monodromies of the electrons themselves are trivial -- the phase factor that results from moving any particle around an electron is unity -- this connection does not give any direct information about the electronic state of the QH fluid. However, it is suggestive that in the CFT language the quasiparticles are represented by local vertex operators, such that the topological properties are coded in the algebraic properties of these operators. The simplest example is the $\nu= 1/m$ Laughlin state, where a quasihole wave function takes the form of a correlator of the chiral vertex operator $H=\rme^{\rmi \varphi /\sqrt m}$, where $\varphi$ is a massless chiral scalar field, compactified on a circle of radius $\sqrt m$. It is then tempting to assume that the electrons can be represented by charged vertex operators with trivial monodromies, the simplest of which in this example is $V_1=\rme^{\rmi \sqrt m \varphi}$. In fact, the $\nu = 1/m$ Laughlin wave function can be written as
\be{lau}
\Psi_{1/m}(z_1 \dots z_N) =\prod_{i<j}^N (z_i - z_j)^m   \rme^{-\sum_{i}^N |z_i|^2/4\ell^2} =  \av{ \prod_{i}^N V_1(z_i){\cal O}_{bg}  } \, ,
\ee
where ${\cal O}_{bg}$ denotes a background charge, needed to define the correlator\cite{mr}; the issue of background charge will be discussed further in section \ref{sec:bgc}.

The other  line of reasoning relating Chern-Simons theory to CFT concerns the gapless edge modes that characterize the quantum Hall fluids. Wen has given several arguments for why these edge modes should be described by chiral bosons, or a 1+1 dimensional CFT\cite{wen}. In particular, he stressed that in the presence of boundaries, a Chern-Simons theory is no longer purely topological, but supports such gapless edge modes with non-universal velocities. Technically, these edge modes are necessary to preserve the gauge invariance of the Chern-Simons theory, which is broken by the presence of boundaries. This gauge anomaly, which would  destroy current conservation, is cancelled by the chiral anomaly of the edge theory, as first discussed in this context by Stone\cite{stone}. Physically this anomaly cancellation amounts to the conserved charge being transferred from the bulk to the edge. This imposes strong constraints on the CFT describing the edge physics, since the edge must support excitations with the same fractional charge as the bulk. It is thus quite natural to assume that {\em both the dynamics of edge excitations, and the bulk wave functions of electrons and quasiparticles, are described by the same CFT}. This conjecture, which was first formulated by Moore and Read, has been proven to be very fruitful in the study of QH liquids, since it has allowed for the construction of model wave functions even in cases where the Chern-Simons theory is not known. 


\subsection{The hierarchy of Abelian QH states}   
\noi
It should be emphasized  that the QH-CFT connection does not tell us {\em which} CFT should be associated with a particular observed QH state, and thus does not provide any immediate prediction of ground state trial wave functions or quasiparticle quantum numbers. One way to proceed is by brilliant guess work in the spirit of Laughlin; this was how the famous Pfaffian, or Moore-Read state, believed to describe the $\nu = 5/2$ state observed in the second Landau level, was found\cite{mr}. There is, however, a large class of QH states for which there exists a more systematic approach, namely, the fully polarized states in the lowest Landau level, which are the main subject of this paper. At a qualitative level, all observed states in this class can be understood in the framework of the Haldane - Halperin hierarchy picture\cite{halphi,haldhi}. The idea is that as the magnetic field is tuned away from the center of a given QH plateau, more and more quasiparticles are created. At a certain density of these quasiparticles, they will form a strongly correlated state of the Laughlin type, leading to a new, homogeneous "daughter" QH state at a different filling fraction. Using some quite natural assumptions, one can then determine the charge and fractional (Abelian) statistics of the excitations in this new QH state. How many such daughter states one is able to observe depends on experimental details, the limiting factor usually being the mobility of the sample. It is natural to assume that states with the highest gap $\Delta_{qp,qh}$ to quasiparticle excitations will be the most prominent. There are strong reasons to believe that this gap decreases monotonically as $1/q$ decreases, where $\nu = p/q$ is the filling fraction\cite{bekar,bergh}. This expectation is strongly supported by experiments. 

Even though the idea of a hierarchy can explain why one observes only odd integer fractions (in fact {\em all} such fractions occur in the hierarchy) and also roughly the order in which they are observed, it does not by itself provide a complete and satisfactory understanding of the abelian QH states. One reason is that the wave functions given in the original papers were rather complicated, and there has been only limited progress in evaluating them numerically.\footnote{
There are in fact slightly different versions of the the hierarchy wave functions. As far as we know, the ones proposed by Haldane have been tested the most exhaustively, albeit for very small systems\cite{greiter}.}  
As mentioned in the introduction, this issue was partially resolved in a series of recent papers, which present a CFT description of all hierarchy states obtained by successive condensation of quasi\emph{electrons}\cite{hansson07,bergh,hhv08,hhv09}. 

\subsection{Conformal spin, orbital spin and the shift} \label{shift}
\noi
Another, and more important, reason for why the idea of a  hierarchy per se does not provide a fully satisfactory description of the QH states  is that it does not predict the shift of the various wave functions.  The shift $S$ of a QH fluid describes its response to curvature, and can be extracted from the ground state wave function. It can be defined by
\be{shiftex}
N_\Phi = \frac 1 \nu N - S \, ,
\ee
where $N_\Phi$ is the number of flux quanta penetrating the system. On flat geometry, such as torus, the shift always vanishes, while on the sphere it is a nontrivial characteristic of the state. A simple example is that of $n$ completely filled  Landau levels. The degeneracy of the $n^{th}$ Landau level on the sphere is $N_{\Phi} + (2n-1)$, so for $\nu=n$ one finds $N = nN_{\Phi} + n^2$, and thus the shift $S = n$. A $\nu = 1$ state with only the  $n^{th}$ Landau level filled, would have a shift $S = 2n-1$. This result can be understood as a consequence of the electrons carrying an orbital spin $S_o = n-1$ in the $n^{th} $ Landau level, and recalling that rotating a particle with  spin on a curved surface gives rise to a Berry phase\cite{wen}. In the context of CFT one can derive a simple general relation -- the shift on the sphere is twice the average conformal spin of the constituent particles\cite{read08}. In our approach to the hierarchy, the conformal spins of all operators are determined by how the various condensates are formed, which gives a precise prediction for the shift.

\subsection{The hierarchy and the Wen classification.}
\noi
A different route to describing the hierarchy states has been taken by Wen and collaborators. In a series of papers\cite{wenclass}, they have given a classification of all Abelian QH liquids based on an effective multi-component Chern-Simons theory. The filling fraction and topological data, \ie the charge and statistics of all the quasiparticles, as well as the shift,  can be calculated from three objects -- the K-matrix, $K_{\a\b}$, the spin vectors, $s^{(\alpha)}$, and the $t$-vectors, $t^{(\alpha)}$. Under some assumptions, it is possible to deduce the explicit form of these objects for any Abelian hierarchical state; fixing the spin vectors amounts to postulating the orbital spin of the constituent particles, while, in general, the rank of the K-matrix matches the level of the hierarchy, and the signatures of its eigenvalues directly translate to the number of chiral and anti-chiral edge modes. This will be important in the following, as a CFT description of the Abelian hierarchy should be consistent with this classification; we will see explicit examples of how this works in the following sections. 

\section{ Hierarchy wave functions from CFT} \label{cfthier}
\noi
In order to set the stage for what follows, we here briefly review the CFT approach to the Abelian hierarchical states that result from successive condensation of quasielectrons, a subset of which is the positive Jain series\cite{bergh,hhv08,hhv09}. The relationship between the hierarchy and composite fermion (CF) wave functions is a controversial question that has proven hard to settle. For instance, Read has argued that the composite fermion wave functions do form a hierarchy\cite{read90}, and in Ref. \onlinecite{bekar} Bergholtz and Karlhede showed that all Jain states are explicitly  hierarchical when evaluated on a thin cylinder. On the other hand Jain claims  that the composite fermion wave functions are qualitatively different from the proposed hierarchical states.\footnote{See chapter 12 in Ref. \onlinecite{jainbook}.} 

This debate was partially settled in Refs. \onlinecite{hhv08} and \onlinecite{hhv09}, where it was shown explicitly that the composite fermion wave functions in the positive Jain series,  $\nu = n/(2pn + 1)$, can be exactly obtained by sequential condensations of quasielectrons. This result is based on the construction of  a quasi-local quasielectron operator  ${\cal P}(\bar\eta)$, such that a state with $M$ quasielectrons at the positions $\bar \eta_i$  is obtained by insertions of this operator into the ground state correlator, in precise analogy to what is done to obtain quasihole states.  ${\cal P}(\bar\eta)$ creates the excess local charge corresponding to a quasielectron  at $\bar \eta$, by {\em shrinking the correlation holes} around the electrons in the vicinity of this point. For example, a state of $M$ quasielectrons in the Laughlin state \pref{lau} can be written as
\be{altqe}
\Psi_{1/m}(\bar\eta_1\dots \bar\eta_M ; z_1 \dots z_N) =  \av{  \prod_j^M {\cal P}(\bar\eta_j) \prod_i^N V_1(z_i) {\cal O}_{bg} } \, ,
\ee
just as a multi-quasihole state is obtained by multiple insertions of the hole operator $H(\eta_i)$.  Explicit hierarchy wave functions can then be obtained by convoluting multi-quasielectron states of the type \pref{altqe} with appropriate pseudo wave functions
\be{pseu}
\Phi^{\star}_k = \prod_{i<j}^M (\eta_i - \eta_j)^{k}  \rme^{-\sum_{i=1}^M |\eta_i|^2/4m\ell^2}  \, ,
\ee 
which describe the quasielectron condensates. The integer $k$ has to be such that the electronic wave function does not vanish under antisymmetrization, though different $k$ give different states in the hierarchy. Successive condensation of quasielectrons then yields wave functions of the form
\be{genstate}
\Psi  &=& {\cal  A}  \langle\prod _{\alpha=1} ^{n}\prod_{i\in M_\alpha} ^{|M_{\alpha}|}V_\alpha (z_i)  {\cal O}_{bg} \rangle    \nonumber \\
 &=& {\cal A} \{ (1-1)^{ K_{11} } \partial_2 (2-2)^{K_{22}}\dots  \partial_n^{n-1}(n-n)^{K_{nn}}  \\
&\times&  (1-2)^{K_{12}} (1-3)^{K_{13}} \dots ((n-1)-n)^{K_{n-1,n}} \} \,\rme^{-\sum_{i}^N |z_i|^2/4\ell^2}\ \nonumber  \
\ee
that can be thought of as hierarchical generalizations of the Laughlin wave function \pref{lau}. Here $K_{\a\b}$ are the elements of the corresponding K-matrix in Wen's classification, and the vertex operators describing the electrons are given by $V_\alpha (z) = \partial_z^{\a-1} \rme^{\rmi\vec Q_\a\cdot\vec \varphi (z)}$ such that $\vec Q_\alpha \cdot \vec Q_\beta=K_{\alpha\beta}$, and $\vec \varphi$ is a multiplet of compactified scalars.\footnote{
Throughout this paper, we assume a diagonal action for the field $\vec\varphi$ so that 
$\langle \varphi_a(z) \varphi_b (w) \rangle =- \delta_{ab}\ln (z-w)$.}
We have introduced the notation
\be{notation}
\partial_\a^k (\a-\a)^{K_{\a\a}} &\equiv&  \prod_{i\in M_\a}^{|M_\a|} \partial_{z_i}^k 
\prod_{i<j\in M_\a}(z_i-z_j)^{K_{\a\a}}  \nonumber
\\
(\a-\b)^{K_{\a \b}} &\equiv&  \prod_{i\in M_\a}^{|M_\a|} \prod_{j\in M_\b}^{|M_\b|} (z_{i}-z_{j})^{K_{\a \b}} \ ,
\ee
where $z_{i}$ with $i\in M_\a$ are the $|M_\a|$ coordinates associated with the vertex operator $V_\a$, and the ordering is such that the derivatives in \pref{notation} are at the very left in \pref{genstate}. We suppress the antisymmetrizer $\cal A$ and the common gaussian factor in the following.

Although the wave functions \pref{genstate} uniquely follow from the above method to form condensates, the vectors $\vec Q_\a$, and thus the electron operators, are determined only up to a similarity transformation that leaves $K_{\a\b}$ unchanged. We exemplify this with the simplest case -- a maximally dense condensate in the $\nu = 1/3$ Laughlin state. Performing the condensation by convoluting \pref{altqe} with \pref{pseu}, and taking $k=0$, we obtain the  $\nu = 2/5$ state $\Psi_{2/5} = (1-1)^3 \partial_2 (2-2)^3 (1-2)^2$ ($k=2$ gives $\nu = 4/11$), which is of the form \pref{genstate} with 
\be{v2}
V_1(z) &=& \rme^{\rmi\sqrt 3 \vphi_1 (z) } \\
V_2(z)   &=&  \partial_z \rme^{ \rmi \frac 2 { \sqrt{3} } \vphi_1 (z) + \rmi\sqrt{\frac 5 3} \vphi_2 (z) } \nonumber \, .
\ee
As discussed in Ref. \onlinecite{torus}, there is an alternative choice of the vectors $\vec Q_\a$,
\be{v3}
\tilde V_1(z) &=& \rme^{\rmi\sqrt {\frac 5 2} \phi (z) + \rmi \sqrt {\frac 1 2  }\vphi (z) } \\
\tilde V_2(z)   &=&  \partial_z \rme^{\rmi\sqrt {\frac 5 2} \phi (z) - \rmi \sqrt {\frac 1 2  }\vphi (z) }  \nonumber \, .
\ee
which corresponds to a decomposition of the K-matrix into a charged part described by the field $\phi$, and a neutral, topological,  part, described by $\vphi$. Such decomposition into a charged and a topological part can always be made, and has the advantage that only one field needs a neutralizing background (recall that $\tilde V_1$ and $\tilde V_2$ appear in equal numbers in the correlator). It thus incorporates the physically motivated expectation that there is only one gapless charged edge mode. From a theoretical  point of view this representation is appealing since it can be used to motivate a plasma analogy\footnote{
This point has been stressed by Read in connection with the non-Abelian Pfaffian states, see Ref. \onlinecite{read08}. },
while on the negative side,  it has no manifest hierarchical structure.


\section{Anti-holomorphic blocks in QH wave functions } \label{sec:antihol}

\noi
The electric charges of the electrons and quasiparticles are closely related to the charges of the corresponding vertex operators under the $U(1)$ transformations $\varphi_\alpha \to \varphi_\alpha + a_\alpha$. A positive charge particle, \ie a quasihole,  is represented by a vertex operator $H(\eta) = \rme^{\rmi\alpha \vphi (\eta)}$ with $\alpha > 0$, and the polynomial part of the (conjugated) quasihole pseudo wave function $\Phi^{\star}$ is thus anti-holomorphic in the $\eta$:s.\footnote{
Note that in this picture the $U(1)$ charge of the electron operator is the same as that of a hole. This is because the $U(1)$ charge is related to the vorticity, \ie the depletion or "contraction" of the electron fluid. Thus a physical electron is viewed as a point-like negative charge, together with a positively charged vortex.This combination is neutral, which correctly reflects the fact that an electron can be added locally to the liquid without changing the local charge. The only effect is that the whole system expands slightly. This picture which is originally due to Read\cite{read89} is discussed in some detail in Ref. \onlinecite{hansson07}. }
This is in contrast to the quasielectron case, where the operator ${\cal P}(\bar\eta)$  depends on the anti-holomorphic coordinates $\bar\eta_i$, and a pseudo wave function such as \pref{pseu} depends on the holomorphic $\eta_i$. If we further recall\cite{hhv09} that the holomorphic Jastrow factor $ \prod_{i<j}^M (\eta_i - \eta_j)^{k} $ in \pref{pseu} gives rise to the corresponding factor $ \prod_{i<j}^M (z_i - z_j)^{k} $ in the electronic wave function, we would, in analogy, expect hierarchical wave functions describing quasihole condensates to involve {\em anti-holomorphic Jastrow factors}. This conclusion is not only consistent with the expected form of the wave function of positively charged particles in a magnetic field, but is also supported by the demand that the chirality of the edge modes of quasihole condensates is opposite to those supported by quasielectron condensates. In the framework of conformal field theory, it is then natural to consider wave functions which involve products of holomorphic and anti-holomorphic conformal blocks.  We now explain how to extract LLL wave functions from such products.


\subsection{A coherent state representation}
\label{sec:cs}

\noi
As a  first example we revisit the Laughlin state and in that context introduce a coherent state formalism that will be important in the following. As a possible modification of Laughlin's wave function at $\nu = 1/m$, Girvin and Jach\cite{girvinjach} proposed the following expression:
\be{girjach}
\tilde\Psi_{1/3} (z_1\dots z_N) &=& \int \frac{d^2\xi}{2\pi} \, \rme^{\sum_i (-\frac 1 4 |\xi_i|^2 + \half \bar\xi_i z_i -\frac 1 4 |z_i|^2)} \prod_{i<j}^N (\xib_i -\xib_j)^k  \prod_{i<j}^N (\xi_i -\xi_j)^{m+k}\rme^{-\sum_i \frac 1 4 |\xi_i|^2},
\ee
where we have set $\ell=1$. This formula can be thought of in two closely related ways. The first is to notice that the integrand contains as a factor a LLL coherent state (CS) 
 \be{cs}
\rme^{\sum_i (-\frac 1 4 |\xi_i|^2 +\half \bar\xi_i z_i -\frac 1 4 |z_i|^2)}  = \bracket{z_1\dots z_N}{\xib_1\dots\xib_N},
 \ee
which describes $N$ particles maximally localized to the points  $\{\xi_i\}$. This wave function is normalized with respect to the measure $\frac{\rmd^2\xi}{2\pi}$. The remaining factor in \pref{girjach} is a wave function for $N$ charged particles, with coordinates $\xi_i $ and $ \bar\xi_i$, in a magnetic field. It can also be thought of as the coherent state wave function of the state defined by the left-hand-side of \eqref{girjach}. We will see that, in general, it is this wave function that is directly related to correlators in CFT, not its LLL representation.

Alternatively, one can note that for each coordinate label $z_i$ and $\xi_i$ the coherent state \eqref{cs} is proportional to the LLL representation of the delta function
\be{llldel}
\delta_{\textrm{LLL}} (z, \xi) = \frac{1}{2\pi}\rme^{-\frac 1 4 |z|^2 + \half\bar\xi z - \frac 1 4 |\xi|^2},
\ee
which allows us to reverse the above logic and think of the LHS \eqref{girjach} as the LLL projection of the coherent state wave function that need not be holomorphic. For integer $k$ this projection is easy to carry out, and one obtains
\be{girjach2}
\tilde\Psi_{1/3}  (z_1\dots z_N) \propto \rme^{-\frac 1 4 \sum_i|z_i|^2 } \prod_{i<j}^N (\partial_{z_i}  - \partial_{z_j}  )^k  \prod_{i<j}^N (z_i -z_j)^{m+k}  \, .
\ee
That this expression is a close relative of the Laughlin wave function is seen by rewriting it as 
\be{laucor}
\tilde\Psi_{1/3} &=& \int \prod_i^N d^2\xi_i \, \delta_{\textrm{LLL}} (z_i, \xi_i) \,  
\prod_{i<j}^N |\xib_i -\xib_j|^{2k}  \prod_{i<j}^N (\xi_i -\xi_j)^{m} 
\rme^{-\frac{1}{4}\sum_i^N |\xi_i|^2},
\ee
which is a LLL projection of the Laughlin wave function multiplied with additional correlation factors between all pairs of particles.

We  now rewrite \pref{girjach} in terms of CFT correlators. Introducing  the following product of a chiral and an anti-chiral vertex operator:
\be{girop}
V_{(m,k)} (\xi,\xib) \equiv \rme^{\rmi\sqrt{m+k}\varphi(\xi) } \rme^{+ \rmi\sqrt{k}\bar\varphi(\xib)},
\ee
we can write
\be{girjach3}
\tilde\Psi_{1/3}  (z_1\dots z_N) &=&  \int \frac{d^2\xi_i}{2\pi}
 \bracket{z_1\dots z_N}{\xib_1\dots\xib_N}  \, 
 \av{ \prod_i^N V_{(m,k)}(\xi_i,\xib_i){\cal O}_{bg}  } \nonumber \\
  &=& \av{ \prod_i^N {\cal V}_{m,k} (z_i){\cal O}_{bg}  },
\ee
where
\be{qlocel}
{\cal V}_{m,k} (z)= \int \frac {d^2 \xi} {2\pi}\, \rme^{-\frac { |\xi|^2}  4  + \half \bar \xi z -\frac{|z|^2} 4 }\,  V_{(m,k)} (\xi,\xib) .
\ee
The two lines in \pref{girjach3} illustrate an important point. The wave function can be viewed alternatively as a LLL projection of products of holomorphic and anti-holomorphic conformal blocks, or as a CFT correlator of the {\em quasi-local electron operators} ${\cal V}_{m,k} (z)$.\footnote{
Note that, up to a phase, the exponential in \pref{qlocel} is just the gaussian factor $ e^{-|\xi - z|^2/4\ell^2}$, where we have restored the magnetic length. }
 The latter interpretation is very appealing for two reasons. First, it incorporates the notion that particles in the LLL are inherently fuzzy in the sense that they cannot be localized more precisely than the magnetic length. Second, it puts electrons and quasiparticles on the same footing. We have already stressed that the natural CFT description of quasielectrons is by means of a quasi-local operator ${\cal P}(\bar\eta)$, and one can also introduce a similar quasi-local operator ${\cal H}(\eta)$ for the quasiholes\cite{coming}. When explicitly building the hierarchy by constructing quasihole condensates, it is in fact the operators ${\cal H}(\eta_i)$, rather than the local $H(\eta_i)$, that must be used in order to find a quasihole counterpart to the quasielectrons state \pref{altqe}. Only in this way will the integrals over the quasihole positions become tractable\cite{coming}. It is in fact  only in the simplest cases, such as the Laughlin states or the CF states in the positive Jain sequence, that holes and electrons are represented by fully local operators. In the following we will, for practical purposes, mainly use the language of LLL projection, but we stress that the interpretation in terms of non-local electron and quasiparticle operators is  more natural.


\subsection{The background charge} \label{sec:bgc}

\noi
In deriving \pref{girjach3}, one must be careful in treating the neutralizing background charge. We consider the holomorphic and anti-holomorphic sectors as independent, which requires that to define the correlators in \pref{girjach3} independently background fields must be included for both $\varphi(\xi)$ and $\bar\varphi(\xib)$. A direct calculation, using the methods explained in Refs.  \onlinecite{hhv08} and \onlinecite{hhv09}, yields exponential factors $\exp (-(m+k)|\xi_i|^2/4m\ell^2)\exp (-k |\xib_i|^2/4m\ell^2)$ rather than $\exp (-|\xi_i|^2/4\ell^2)$ in \pref{girjach}. The former is precisely what is expected from two sets of particles with charges $(1+\frac k m)e$ and $- \frac k m e$ moving in the magnetic field $eB = 1/\ell^2$, while the latter would be appropriate for the center of mass motion of the charge $e$ composite of these particles. In this case, the charges are easily understood, since the Jastrow factor $ \prod_{i<j}^N (z_i -z_j)^{m+k} $ in \pref{girjach2} is precisely what one gets by putting $k$ charge $e/m$ holes on top of each electron. The anti-holomorphic Jastrow factor amounts to attaching equally many quasielectrons to get back the charge $e$, albeit with a different microscopic wave function. 

The lesson here is that for the product of  correlators in \pref{girjach3} to be interpreted as describing  $N$ particles, rather than $2N$, we must give a prescription for how to identify $\xib$ and $\xi^\star$. The most natural way is to introduce relative and center of mass coordinates $\xi_r = \xi - \xib^\star$ and $2\xi_{cm} = \xi + \xib^\star$, and identify $\xib=\xi^\star\equiv \xi_{cm}^\star$ in the limit $\xi_r\rightarrow 0$. The exponential factor for the CM part then indeed becomes  $\exp (-|\xi_i|^2/4\ell^2)$, as explained in Appendix \ref{magn}. While this example might appear a bit contrived -- why should the electron in the Laughlin state be though of as composed of the bare electron superimposed with an equal number of quasiholes and quasielectrons -- we will see in the next section that this provides a natural language for the study of hierarchical states and their relation to CFT correlators. In the following we will, without any further ado, always use the gaussian factor appropriate to the center-of-mass coordinate when fusing holomorphic and anti-holomorphic operators.


\section{ A two-component picture of hierarchical states } \label{twocomp}

\noi
In this section we will construct wave functions for three hierarchical states that involve quasihole condensates. The first is the $\nu = 2/3$ state, which has been discussed extensively in the literature. It is the particle-hole conjugate of the Laughlin $\nu = 1/3$ state but can also be described as a "reversed-flux attachment" CF state\cite{wu93}. We will view it as a hole condensate in the first filled Landau level, and use the CFT formalism to construct trial wave functions. We then turn to two "mixed" states, which contain both quasielectron and quasihole condensates.  We  will see that the states can be described by a two-component picture, with each component a bosonic or fermionic QH state, one with positive and the other with negative charge.  Being Abelian QH states, each of the components is in turn built from a number of strongly correlated components. In the CF picture, these components correspond to composite fermions in different effective Landau levels, while in our CFT formalism they are formed by the nonequivalent electron operators. The two components are correlated by matching the coordinates $\xi_i$ and $\bar\xi_i$ in the different groups, and identifying the center-of-mass coordinate as discussed in the previous section. Clearly, this matching puts strong restrictions on the two components since both number, charges and statistics of the particles  must combine so as to describe a fermionic state of unit charge particles as is appropriate for an electron system. The principles illustrated by the following examples apply to general Abelian hierarchy states; a more formal discussion will be given in Ref. \onlinecite{coming}.

\subsection{The $\nu = 2/3$ state} \label{23state}
\noi
We start by studying the very simplest quasihole condensate state, namely that at $\nu = 2/3$. The $\mathrm K$-matrix is given by\cite{wen}
\be{kmat}
\mathrm K = \left(
\begin{array} {cc}
1 & 2 \\
2 & 1 
\end{array} \right).
\ee
Following \pref{genstate} we would thus expect a wave function of the form 
\be{wrong}
  (1-1) (2-2) (1-2)^2\, ,
\ee
but it is easy to show that since K has a negative eigenvalue, \pref{wrong} cannot be written as a correlator of purely holomorphic vertex operators of the type $\rme^{ \rmi Q^{\alpha}_i \varphi_i }$ with $\alpha = 1,2$. Instead, we decompose the K-matrix into two positive definite components, $\mathrm K = K - \bar K$. One such decomposition is
\be{kmatde}
\mathrm K = K - \bar K = 
 \left(\begin{array} {cc}
3 & 2 \\
2 & 3 
\end{array} \right) - 
 \left(\begin{array} {cc}
2 & 0 \\
0 & 2 
\end{array} \right) ,
\ee
so that $K$ describes a $\nu = 2/5$ Jain  state of charge $5e/3$ particles, while $\bar K$ describes a bosonic $\bar\nu=1$ quantum Hall state made up of two uncorrelated layers of equally many charge $-2e/3$ particles. Computing the corresponding filling factors of the positive and negative charges\cite{wen}, we note that they are combined as $1/\nu - 1/\bar\nu = 3/2 = 1/\nu_{tot}$;  this is in fact a general result\cite{coming}. The K-matrix now suggests the wave function
\be{wf23}
\Psi_{2/3}  = (\bar 1 - \bar 1)^2 \bar\partial_2 (\bar 2 - \bar 2)^2 (1-1)^3  (2-2)^3 (1-2)^2 \, ,
\ee
where we have extended  the previous shorthand notation \pref{notation} by
\be{aholprod} 
\bar\partial_\a^{k} (\bar\a- \bar\a)^{\bar K_{\a\a}} &\equiv&  \prod_{i\in \bar M_\a}^{|\bar M_\a|} {z^k_{i}} 
\prod_{i<j\in \bar M_\a}(\partial_{z_i} - \partial_{z_j} )^{\bar K_{\a\a}}  \\
(\bar\a-\bar\b)^{\bar K_{\a \b}} &\equiv&  \prod_{i\in \bar M_\alpha}^{|\bar M_\a|} \prod_{i\in \bar M_\b}^{|\bar M_\b|} (\partial_{z_i} -\partial_{z_i} )^{\bar K_{\a \b}} \ . \nonumber
\ee
Here, $z$ and $\partial$ are the LLL projected versions of $\bar\partial$ and $\bar z$, respectively. This wave function can be obtained by a coherent state projection, as described in Section \ref{sec:antihol}. The attentive reader has noticed that we have introduced a product of derivatives $\bar\partial_2$ in the wave function \pref{wf23}, that is not implied by the K-matrix alone. The origin of this is analogous to that of the holomorphic derivatives in the general expression \pref{genstate} for states formed by quasielectron condensation\cite{coming}. Since $K$ and $\bar K$  in the decomposition \pref{kmatde} are both positive definite, and define level-two hierarchical states with $|M_i| = |\bar M_i|$, the chiral and anti-chiral vertex operators can be combined into two full electron operators
\be{felop}
V_1 &=&  \rme^{\rmi\sqrt 3 \vphi_1 +\rmi \sqrt 2 \vphib_1} \\
V_2 &=& \bar \partial \rme^{ \rmi\frac 2 {\sqrt 3} \vphi_1 +\rmi \sqrt{\frac 5 3} \vphi_2 + i\sqrt 2 \vphib_2 } \, . \nonumber
\ee
In analogy with \pref{girop} and \pref{girjach3}, the  wave function can now be written as 
\be{wf23b}
\Psi_{2/3} = {\cal A} \left[ \{\bar\xi \}, \{ z \} \right] 
 \av{ \prod_{i\in {M_1}} V_{1}(\xi_i,\xib_i)      \prod_{j \in {M_2}}  V_{2}(\xi_j,\xib_j)      {\cal O}_{bg}  }, 
\ee
where the coherent state projection kernel is defined by
\be{kerder}
{\cal A}\left[ \{\bar\xi \}, \{ z \} \right] = {\cal A} \left[ \int \prod_i^N\frac {d^2\xi_i} {2\pi} \bracket{z_1\dots z_N}{\xib_1\dots\xib_N} \right].
\ee
Alternatively, we can introduce the non-local electron operators ${\cal V}_i$ in direct analogy with \pref{qlocel} and write,
\be{wf23b1}
\Psi_{2/3} = {\cal A}  \av{ \prod_{i\in {M_1}} {\cal V}_{1}(z_i)     \prod_{j \in {M_2}}  {\cal V}_{2}(z_j)       {\cal O}_{bg}  } \, .
\ee
Such alternative expressions in terms of quasi-local electron operators can be given for a general state in the hierarchy. Although for simplicity, we have not presented the explicit hierarchical construction of the wave function\cite{coming}, it is explicitly of hierarchical form since the correlator $ \av{ \prod_{i\in {N}} {\cal V}_{1}(z_i)   }$ equals the wave function for a filled LLL. 

The decomposition \pref{kmatde} of the $K$-matrix is not unique. For example, both $K$ and $\bar K$ are positive-definite for all integer-valued decompositions of the form
\be{de}
\mathrm K = K - \bar K =  \left(\begin{array} {cc}
1+k & 2+l \\
2+l & 1+k 
\end{array} \right) - 
 \left(\begin{array} {cc}
k & l \\
l & k 
\end{array} \right) 
\ee
if $k > l+1$, and singular if $k=l+1$.  Also $|M_i| = |\bar M_i|$, and it can be checked explicitly that the filling fractions again combine as $1/\nu - 1/\bar\nu = 3/2 = 1/\nu_{tot}$. 

To directly relate our scheme to the CF picture, consider the singular case
\be{kmatde-a}
\mathrm K = K - \bar K = 
 \left(\begin{array} {cc}
2 & 2 \\
2 & 2 
\end{array} \right) - 
 \left(\begin{array} {cc}
1 & 0 \\
0 & 1 
\end{array} \right) \, ,
\ee
where $K$ is of rank 1. This implies that the holomorphic part of the wave function can equivalently be described by the one-dimensional positive-definite matrix $K=2$. Hence, a natural basis for the vertex operators is
\be{felop2}
\tilde V_1 &=&  \rme^{\rmi\sqrt 2 \vphi +\rmi\vphib_1}    \\
\tilde V_2 &=& \bar \partial \rme^{\rmi\sqrt 2 \vphi +\rmi\vphib_2}    \, . \nonumber
\ee
A correlator with insertions of these operators is the Jain $\nu = 2/3$ "reverse flux attachment" wave function\cite{wu93}
\be{wf23jain}
\Psi_{2/3}^{\mathrm{Jain}}  = (\bar 1 - \bar 1)\bar\partial_2 (\bar 2 - \bar 2) (1-1)^2  (2-2)^2 (1-2)^2,
\ee
which, when formulated on a sphere, is numerically very close to the Coulomb wave function, as well as to the exact particle-hole conjugate of the Laughlin 1/3 state. The wave function \pref{wf23b} remains to be tested numerically. 

In our discussion of the holomorphic hierarchical states, we pointed out that there is always a representation of the vertex operators where only one field is charged, and thus requires a background. Such a representation can also be found in the more general case including quasihole condensates. Again, it amounts to decomposition of the $\mathrm K$-matrix into a charged and a topological part, which in this case gives
\be{kmatde-m}
\mathrm K = K_c + K_{top} = \frac 3 2 
 \left(\begin{array} {cc}
1 & 1 \\
1 & 1 
\end{array} \right) + \half 
 \left(\begin{array} {cc}
-1 & 1 \\
1 & -1 
\end{array} \right) \, .
\ee
That the topological K-matrix has rank 1 immediately suggests the following parametrization of the vertex operators:
\be{23min}
  V_1^{\mathrm{min}} &= \rme^{\rmi \phi_c/\sqrt{\nu}}\rme^{\rmi\bar\phi_c/\sqrt{2}}\\
  V_2^{\mathrm{min}} &= \bar\partial \rme^{\rmi \phi_c/\sqrt{\nu}}\rme^{-\rmi\bar\phi_c/\sqrt{2}}   \nonumber \, ,
\ee
so that the corresponding wave function reads
\be{wf23one}
\Psi_{2/3}^{\mathrm{min}}  = (\bar 1 - \bar 1)^\half \bar\partial_2 (\bar 2 - \bar 2)^\half  (\bar 1 - \bar 2)^{-\half} \prod_{i<j} (z_i-z_j)^{\frac{3}{2}} \, .
\ee
Note that in spite of the fractional powers, this expression is single-valued in all the coordinates, so the  coherent state projection is well defined and gives a unique, although rather complicated, LLL wave function.  Also note that this wave function differs from the Jain state \pref{wf23jain} only by the real  Jastrow  factor  $\prod_{i<j} |z_i-z_j|$, which only adds an equal extra repulsion between all the electrons. We might even speculate that from the CFT viewpoint the minimal state \pref{wf23one} is more fundamental than the Jain state \pref{wf23jain}, the two being related just as  the Laughlin state \pref{lau} to the  Girvin-Jach state \pref{girjach}.

Additionally, we note a close analogy to the description of the holomorphic state at $\nu=2/5$, using the operators \pref{v3}, in which the following decomposition of the $K$-matrix is manifest:
\be{}
\mathrm K = K_c + K_{top} = \frac 5 2 
 \left(\begin{array} {cc}
1 & 1 \\
1 & 1 
\end{array} \right) + \half 
 \left(\begin{array} {cc}
1 & -1 \\
-1 & 1 
\end{array} \right) \, .
\ee
The corresponding wave function is given by
\begin{equation}
  \Psi_{2/5}= 
  (1-1)^\half \partial_2 (2-2)^\half (1-2)^{-\half} \prod_{i<j} (z_i-z_j)^{\frac{5}{2}} \, ,
\end{equation}
which suggests that the neutral sectors of the hierarchy states at $\nu=2/3$ and $\nu=2/5$ differ only in chirality.

To conclude the discussion of the  $\nu = 2/3$  state, we return to the naive guess \pref{wrong}, which we discarded since it does not have a natural expression in terms of CFT correlators. This is not a physical argument, so we must ask whether \pref{wrong} is in some sense pathological. A numerical calculation on a finite system shows that its overlap with the CF wave function on the disk is practically zero for already $N=10$. While this is not enough reason to exclude it, there is a stronger argument based on adiabatic continuity. The basic idea is to examine the various proposed states on a thin torus or cylinder where the quantum Hall problem is exactly solvable for a general family of electron interactions. The ground state, the so-called TT state, is an exact eigenstate that minimizes the electrostatic repulsion. For  $\nu = p/q$, this state is characterized by a periodically repeated unit cell of length $q$ containing $p$ electrons\cite{bekar}.  In Ref. \onlinecite{bergh} it was shown that all the wave functions of the form \pref{genstate} reduce to the exact solutions in the TT-limit. An analogous computation for wave functions containing anti-holomorphic blocks\cite{coming} shows that for all decompositions of the type \pref{de}, including the two proposed wave functions \pref{wf23} and \pref{wf23jain}, as well as \pref{kmatde-m}, the TT-limit agrees with the exact ground state, while for the naive guess \pref{wrong} it does not. This observation seems to generalize -- states that can consistently be described in a two-component picture will also have the correct TT-limit. A detailed discussion of this point, together with an explicit construction of quasihole condensate states and a systematic exposition of the decomposition into charged and neutral components, will be given in another publication\cite{coming}.

\subsection{ Two mixed states  }
\noi
The above description of $\nu=2/3$ can easily be generalized to all states in the negative, or "reverse flux attachment", Jain sequence which covers the filling fractions $\nu  = n/(2np - 1)$. These states can alternatively be described in terms of successive condensation of quasiholes, starting from a filled Landau level. Moreover, a generalization to quasihole condensates off the negative Jain sequence is rather straightforward. For example, the level-two state at $\nu = 4/5$ is the hole condensate counterpart of the 4/11 state mentioned previously; a possible decomposition of its K-matrix, analogous to \pref{kmatde-a}, is
\be{kmatde-54}
\mathrm K = 
 \left(\begin{array} {cc}
1 & 2 \\
2 & -1 
\end{array} \right) =
 \left(\begin{array} {cc}
2 & 2 \\
2 & 2 
\end{array} \right) - 
 \left(\begin{array} {cc}
1 & 0 \\
0 & 3 
\end{array} \right) \, .
\ee
Most of the states in the hierarchy are of mixed type, by which we mean that they contain both quasihole and quasielectron condensates. Such states are indeed observed in experiments. Simple examples of mixed states are the observed $\nu = 5/13$, which can be interpreted as a quasihole condensate on top of the $\nu  =2/5$ Jain state, and  $\nu = 5/7$, a quasielectron condensate in the $\nu = 2/3$ state. For these states there are no simple explicit candidates based on composite fermions,  and the original hierarchy construction offers only rather involved formal expressions. Our aim here is to use the CFT technology to generate explicit trial wave functions for the mixed states, expressed in terms of CFT correlators and amenable to numerical analysis, so that the topological properties of the state are encoded in the CFT operators, just as in the simpler cases discussed above. 

From the previous section it should be clear that because of the ambiguity in the decomposition of the K-matrix, which is present even in the simplest cases, we can at best hope for a construction that, at a given filling fraction, provides a family of trial wave functions, all sharing the same topological properties.

\subsubsection{The $\nu = 5/13$ state}

\noi
This state has been experimentally observed\cite{pan}. In the hierarchy, it is a quasihole condensate on top of the $\nu =  2/5$ state, characterized by the following K-matrix\cite{wen}
\be{k513}
\mathrm K_{5/13} = 
\left(
\begin{array}{cccc}
 3 &   2 & 2   \\
  2 & 3 & 4 \\
  2 & 4 & 3   
\end{array}
\right) =
\left(
\begin{array}{cccc}
 4 &   2 & 2   \\
  2 & 6 & 4 \\
  2 & 4 & 6   
\end{array}
\right) -
\left(
\begin{array}{cccc}
 1 &   0 & 0   \\
  0 & 3 & 0 \\
  0 & 0 & 3   
\end{array}
\right).
\ee
Here, the decomposition into holomorphic and anti-holomorphic parts has been chosen in the simplest way such as to make both $K$ and $\bar K$ positive definite.  Using the methods of Ref. \onlinecite{bergh}, and the coherent state representation introduced in section \ref{sec:cs}, the corresponding wave functions can be expressed as sums of products of holomorphic and anti-holomorphic blocks.  

$K$ and $\bar K$ are chosen so that the relative filling fractions of the three levels are the same for both components, $|M_i| = \bar |M_i|$, and a simple calculation assuming a homogenous state yields  $|M_1| = 3|M_2| = 3|M_3|$.
Thus, just as in the case of the $\nu=2/3$ state, full electron operators can be formed as products of a chiral and an anti-chiral vertex operators. As usual, the explicit expressions for the vertex operators are basis dependent, an example being 
\be{513op}
V_1  &=&  \rme^{2\rmi\vphi_1 + \rmi \vphib_1}  \nonumber \\
V_2 &=&  \partial \rme^{\rmi\vphi_1 + \rmi \sqrt5 \vphi_2 + \sqrt 3 \vphib_2}    \\
V_3 &=&  \bar\partial \partial \rme^{\rmi\vphi_1 + \rmi  \frac 3 {\sqrt 5} \vphi_2 + \rmi \frac 4 { \sqrt 5} \vphi_3 + \rmi \sqrt 3\vphib_3 } \nonumber \, ,
\ee
where $V_1$ and $V_2$ can be used to find a $\nu = 2/5$ wave function in accordance with the hierarchy, \ie they represent the upper left $2\times 2$ submatrices of \pref{k513}. 
Using  the simplified notation \pref{notation}, the corresponding trial wave function can be written as
\be{5/13}
\Psi_{5/13}& =& (\bar 1 - \bar 1) (\bar 2 - \bar 2)^3  \bar\partial_3 (\bar 3 - \bar 3)^3 ( 1 - 1 )^4 \partial_2 ( 2 - 2 )^6 \partial_3 ( 3 - 3 )^6 \\ 
&\times& ( 1 - 2 )^2 ( 1 - 3 )^2 ( 2 - 3 )^4.  \nonumber
\ee
A direct calculation again shows that the TT-limit is correct\cite{coming}. As pointed out, the splitting chosen in \pref{k513} merely corresponds to one example of a large class of trial wave functions that can be constructed for this state, all fulfilling the criteria of correct topological properties and correct TT-limit.  In particular, these include the description in terms of a single charged field, as well as  all splittings where $\bar K$ in \pref{k513} is multiplied by some positive integer. Just as in \pref{laucor}, this corresponds to building in additional correlation factors. Eventually, numerics will have to decide which linear combination of these most closely approximates the Coulomb ground state.

\subsubsection{The $\nu = 5/7$ state}

\noi
This is a quasielectron condensate on top of the $\nu= 2/3$ state. We decompose the K-matrix, which known from Wen's work, as follows
\be{k57}
{\mathrm K}_{5/7} = 
\left(
\begin{array}{cccc}
 1 &   2 & 2   \\
  2 & 1 & 0 \\
  2 & 0 & 1   
\end{array}
\right) =
\left(
\begin{array}{cccc}
 3 &   2 & 2   \\
  2 & 5 & 2 \\
  2 & 2 & 5   
\end{array}
\right) -
\left(
\begin{array}{cccc}
 2 &   0 & 0   \\
  0 & 4 & 2 \\
  0 & 2 & 4   
\end{array}
\right).
\ee
This decomposition satisfies the same formal requirements as the previous examples, which in this case implies that the matrix $\bar K$ is no longer diagonal. As a consequence, the wave function which reads
\be{5/7}
\Psi_{5/7}& =& (\bar 1 - \bar 1)^2  \bar\partial_2  (\bar 2 - \bar 2)^4 \bar\partial_3 (\bar 3 - \bar 3)^4   (\bar 2 - \bar 3)^2    ( 1 - 1 )^3 ( 2 - 2 )^5 \partial_3( 3 - 3 )^5  \nonumber  \\
&\times&  ( 1 - 2 )^2 ( 1 - 3 )^2  ( 2 - 3 )^2  ,
\ee
has off-diagonal anti-holomorphic factors. The relevant vertex operators are easily constructed.

Here we should again remind the reader that, as in the previous cases,  the full explicit wave functions \pref{5/13} and \pref{5/7} do not follow even when we have chosen a particular decomposition of the K-matrix -- $K$ and $\bar K$ do determine the powers in the various Jastrow factors of coordinates and derivatives, but not the explicit factors of $\partial$ and $\bar\partial$ which are crucial for getting the correct shift. In the composite fermion picture these derivatives originate from the projection from higher effective Landau levels, while in the present CFT formalism they originate in the spins of the vertex operators. These in turn follow from the detailed construction of the quasihole condensates from the multi-quasihole states formed by inserting the operators ${\cal H}(\eta)$ into the various correlators\cite{coming}.  For example, one of the vertex operators describing the $\nu =5/7$ state \pref{k57} is given by 
\be{57V3}
V_3(z) = \p \bar \p \rme^{\rmi \sqrt 2 \phi  + \rmi \sqrt 3 \varphi_3 +   2\rmi \bar \varphi_2  }  \, ,
\ee
which has conformal spin $(2 + 3 -4)/2 + 1 - 1 =1/2$. The shift of the state is then given by twice the average of conformal spins of the electron vertex operators as indicated in section \ref{shift}. The result for the three states we have  considered is
\be{shifts}
S_{2/3} = 0 \ \ \ , \ \ \
S_{5/13} = \frac {17} 5  \ \ \ , \ \ \
S_{5/7} =  \frac  3 5 .
\ee
For the first two of these fillings, the shifts have previously been calculated for composite fermions\cite{jainbook,mandal02}, and the results agree with those found here.


\section{Non-Abelian states}
\label{sec:NA}
\noi
Although the main thrust of this paper concerns the Abelian hierarchy, we comment here on the most straightforward applications of our methods to non-Abelian states.

\subsection{Non-Abelian hierarchical states}

\noi
The most direct application of our work would be to provide closed-form trial wave functions for quasihole condensates and general mixed states that derive from a known non-Abelian state, such as the $\nu = 1/2$ Moore-Read Pfaffian state\cite{mr} or the Read-Rezayi generalizations\cite{RR} thereof. In a recent paper, Bonderson and Slingerland (BS) proposed a non-Abelian hierarchy, formed by combining (bosonic) Abelian hierarchy states with a Pfaffian appearing in the Moore-Read state\cite{BS}; they argue that such states may describe many of the quantum Hall states observed in the second Landau level. They discuss the states in the language of the Haldane-Halperine hierarchy and provide explicit trial wave functions as products of a Pfaffian and bosonic composite fermion states. The most direct application of our methods would be to extend this scheme to a full hierarchy based on the Moore-Read state, including quasielectron- and quasihole condensates away from the Jain sequences, and mixed states. On the other hand, the BS scheme involving positive and negative Jain states at second and third level, seems sufficient to provide trial wave functions for the second Landau level states observed to date. It remains to be seen if future experiments will reveal Hall plateaus at fillings that require combinations of a Pfaffian and more complicated, non-Jain, hierarchy states. 


\subsection{The anti-Pfaffian state}

\noi
Another application of our methods is closely related to the so called anti-Pfaffian state\cite{ sslee,levin} which is by definition the particle-hole conjugate of the Moore-Read Pfaffian state. Let us recall the CFT description of the  Moore-Read state due to Cappelli et al\cite{cappelli}, which was used in Ref. \onlinecite{hhv09}. Instead of using Ising fields, the electron operator can be written as
\be{mrel}
V(z) &=& \rme^{\rmi\sqrt 2 \varphi(z)}  \cos(\phi(z)),
\ee
and the Moore-Read wave function as
\be{mrcap}
\psi_{1/2} = {\cal S} \left[ (1-1)^2 (2-2)^2\right] \, \prod_{i<j}^N (z_i - z_j),
\ee
with $\cal S$ denoting symmetrization; the first factor is the bosonic $\nu=1$ Pfaffian state. It is now suggestive that in our scheme we can form another  $\nu=1/2 $ state using the operator
\be{apfel}
V(z) &=&  \rme^{\sqrt{3}i \vphi (z) + i\vphib(z)}      \cos(\bar\phi(\bar z)),
\ee
which gives the wave function
\be{apwf}
\bar\psi_{1/2} = {\cal S} \left[ (\bar 1-\bar 1)^2 (\bar 2-\bar 2)^2\right] \, \prod_{i<j}^N (z_i - z_j)^3,
\ee
which was considered by Jolicoeur in the context of composite boson wave functions\cite{joli}. 

Having identified the underlying conformal field theory, we can now further analyze this state by considering the edge modes. Recall that while the $\nu = 2/3$ Jain wave function \pref{wf23jain} is not the exact particle-hole conjugate of the Laughlin $\nu = 1/3$ state, there are strong reasons to believe that they are in the same universality class, an important point being that they support the same edge modes. Similarly, we propose that the wave function \pref{apwf} is in the same universality class as the anti-Pfaffian. As discussed in Ref. \onlinecite{levin}, the edge of the anti-Pfaffian state can be understood by considering a particle-hole conjugate of the configuration (Pfaffian\,$|$\,vacuum\,$|$\,LLL). The (Pf \,$|$\,vac) interface supports two chiral modes, one charged and one majorana, as is implied by \pref{mrel}, given that $\cos \phi$ is a bosonized chiral majorana field.  The interface (vac\,$|$\,LLL) supports an anti-chiral charged mode. Assuming the vacuum segment to be narrow, one can deduce the modes at the interface (anti-Pfaffian\,$|$\,vac): there is one chiral and one anti-chiral scalar mode, as well as an anti-chiral majorana mode. This is exactly what can be read from the operator \pref{apfel}, which motivates our claim that \pref{apwf} is in the universality class of the anti-Pfaffian. It would clearly be of interest to compare \pref{apwf} with the exact particle - hole conjugate of the Moore-Read state.


\subsection {Non-Abelian quasihole condensates}
In a recent paper Hermanns has generalized the condensation procedure described in Section \ref{cfthier} to form condensates of non-Abelian quasielectrons\cite{hermanns10}.  In the simplest case, the daugher of the bosonic Pfaffian state at $\nu = 1$ is a novel non-Abelian state at  $\nu = 4/3$. Interestingly, the quasiparticles of the daugher state obey $su(3)_2$ fusion rules, while those in the Pfaffian state fuse according to $su(2)_2$. As pointed out by Hermanns, her example can be generalized to a full hierarchy of states formed by successive condensation of non-Abelian quasielectrons. Using the methods developed here, it should be straightforward to extend this to a full  genuinely non-Abelian hierarchy.

\section{Summary and open questions}
\label{sec:sum}

In this paper we have demonstrated, mainly by means of examples, how to extend the CFT description of the spin-polarized quantum Hall liquids from pure quasielectron condensates to the full Abelian hierarchy that includes generic quasiparticle condensates. In this framework, the ground state and quasiparticle wave functions of the system are expressed as correlators of conformal operators which encode the topological properties of the state and carry hierarchically generated orbital spins. 

We have argued that a CFT description of these states requires the introduction of anti-chiral fields in the electron operators, upon which LLL wave functions can be obtained as antisymmetrized correlators of quasi-local electron operators, or equivalently as coherent state projections of conformal blocks involving both chiralities. The chiral decomposition of the K-matrix is, in general, not unique.  Consequently, for a given K-matrix, we have presented a class of possible representative wave functions that we believe are all in the same universality class. An important guideline in narrowing down the number of possibilities has been the demand for agreement with the known exact solutions on a thin cylinder. We emphasize that the wave functions thus obtained involve a coherent state projection, which, however, is always of the same form independent of the level in the hierarchy. This should be contrasted to previous proposals, where a new set of integrals is needed at each level. The main obstacle to numerical analysis of our trial wave functions is therefore not the auxiliary integrals, but rather the need to antisymmetrize high order polynomials in many variables. 

Additionally, we have commented on non-Abelian hierarchy states. In particular we have proposed a wave function that is likely to be in the same universality class as the anti-Pfaffian state. 

There are several obvious future directions. Clearly, numerical work is needed to compare the states proposed in this paper to Coulomb ground states. The possibility of "improving" the trial wave functions by taking linear combinations within a given class was briefly discussed in the case of the Laughlin states, and deserves a more systematic study. Another interesting future direction is to study further the generalizations to non-Abelian states already briefly touched upon in this paper. In this context we note that the question of the relation between our proposed state \pref{apwf}, and the anti-Pfaffian state is but a special case of the more general problem of understanding the role and implementation of particle-hole conjugation in the CFT-based wave functions. All the states considered in this work, and its predecessors, are fully spin polarized; incorporating partially polarized and spin-textured states in the CFT framework remains an open question. We believe, however, that some of the methods developed in the context of GLCS theories might be useful\cite{HKL}. We have repeatedly stressed that the composite fermion wave functions are a subset of the ones considered in this paper. The perhaps greatest success of the composite fermion approach is that it provides a simple and intuitive description of the metallic $\nu=1/2$ state in terms of free composite fermions\cite{HLR}. It is an outstanding problem to understand whether or not this state can be described within the CFT framework.

\acknowledgements
We thank Maria Hermanns for discussions, and Anders Karlhede, Steve Simon and Jaakko Nissinen for discussions and helpful comments on the manuscript. This work was supported by the Norwegian and Swedish research councils, and by NordForsk.

\begin{appendix}

\section{Two charged particles in a magnetic field} \label{magn}
The following is a review of a lowest Landau level system of two charged particles. It serves to illustrate that bound states of the constituent charges can be described within the framework of Born-Oppenheimer approximation such that the resulting center-of-mass functions appear in the same gauge as the constituent degrees of freedom. We will first outline some facts about the single-particle system and then proceed to the two-particle case.

The minimal Hamiltonian for a particle with mass $m$ and charge $e$ on a  plane in a transverse magnetic field $B$ is
\begin{equation}
  H =\frac 1 {2m} \vec\Pi^2,\;\;\; \Pi_i = -\rmi \partial_i + e A_i(\vec r).
\end{equation}
In the Coulomb gauge $\partial_i A_i = 0$, so by Hodge decomposition one can write $A_i = \epsilon_{ij} \partial_j \phi$. The magnetic field strength is then given by $B =-\partial^2 \phi$. Introducing the operators $\Pi_\pm = \Pi_y \mp\rmi\sigma \Pi_x$, where $\sigma = \textrm{sgn}(eB)$, which satisfy $[\Pi_-,\Pi_+] = |eB|$, the Hamiltonian can be cast in the form $H = \frac 1 {2m} \Pi_+\Pi_- +\frac{|eB|}{2m}$. For constant $B$, it then follows that the ground states have the energy $E= |eB|/2m$ and wave functions with the structure
\begin{align}
  \psi_- &= f_-(z) \rme^{-e\phi(z,\zb)}, \; eB\leq 0\\
  \psi_+ &= f_+(\zb) \rme^{e\phi(z,\zb)}, \; eB>0,
\end{align}
where $f_\pm$ are holomorphic in their argument. In the symmetric gauge one has $\phi = -B|z|^2/4$.

Next, consider a translation-invariant system of two charges $e_1$ and $e_2$ with masses $m_1$ and $m_2$. Its Hamiltonian is
\begin{equation}
  H = \sum_{a=1}^2 \frac1 {2m_a} \vec \Pi^2_a + V(r),
\end{equation}
where $\vec \Pi_{a} = \vec p_{a} +e_a \vec A(\vec r_a)$ and $r = |\vec r_1-\vec r_2|$. In general, the center-of-mass and relative motions are not fully separable, but rather are coupled through the magnetic dipole moment $\vec \mu =\mu \, \vec r =  \frac{1}{m_1+m_2}(m_1 e_2-m_2e_1) (\vec r_1-\vec r_2)$. However, there exist a class of gauges where the relative motion can be made explicitly independent of the center-of-mass\cite{bruce}. A canonical change of variables: $\vec R =  (m_1 \vec r_1 + m_2 \vec r_2)/M$, $\vec r = \vec r_1-\vec r_2 $, $\vec P =  \vec p_1+ \vec p_2 $ and $\vec p = (m_2 \vec  p_1-m_1  \vec p_2)/M$ with $M=m_1+m_2$, combined with a gauge transformation $H\to UHU^\dagger$ with 
\begin{equation}
  U = \rme^{\rmi\frac{\mu B}{2}   \vec r_1 \times \vec r_2},
\end{equation}
given in the symmetric gauge, brings the Hamiltonian into the form $H = H_{CM}(\mu) + H_r$, where
\begin{equation}
  H_{CM}(\mu) = \frac{1}{2M}\left(P_i + e A_i(\vec R) +  B \epsilon_{ij}\mu_j\right)^2
\end{equation}
and 
\begin{equation}
  H_{r} = \frac{1}{2m}(p_i+e_{eff} A_i(\vec r))^2 + V(r).
\end{equation}
Here, the reduced mass $m=m_1m_2/M$, $e=e_1+e_2$, and $e_{eff} = \frac 1 {M^2} (m_2^2 e_1 + m_1^2 e_2)$. While the form of the above gauge transformation, known as the Power-Zienau-Woolley transformation, is gauge-dependent, the final result is not.

We are interested in the states in the lowest Landau level for the center of mass coordinate. For concreteness, let us assume that $eB\leq 0$ and $e_{eff}B\leq 0$. Then the LLL wave functions are annihilated by the two operators
\be{holcov}
  \Pi_- &= &
  -2 \partial_{\bar Z} - 2 \partial_{\bar Z} [e\phi(Z,\bar Z)+
  \tfrac{\mu B}{2} \bar Z z ]  \\
  \pi_-     &=& 
  -2\partial_{\bar z}-2 \partial_{\bar z}\left(e_{eff}\phi(z,\bar z)
\right)\, , \nonumber
\ee
which, although not independent, still commute since the dependence of the center-of-mass operator $\Pi_-$ on the relative coordinate  $\vec r$ is purely holomorphic. The LLL problem can therefore be fully solved, provided the potential $V$ is parabolic or absent. However, we are primarily interested in the case where the particles are tightly bound, so that the approximation $\vec r_1 \approx \vec r_2 \approx \vec R$ is valid. For this we assume that the confining potential $V(r)$ is very strong so the size of the low lying bound states is much smaller than the magnetic length $\ell$. This justifies the use of the Born-Oppenheimer approximation, which to leading order amounts to neglecting the slow variables, $\vec R$ and $\vec P $ when solving for the fast variables $\vec r$ and $\vec p$. The Born-Oppenhemer Hamiltonian is obtained by taking an average $\av\dots$ with respect to the ground state of the fast variables. Since this is  an eigenstate of angular momentum, the result is simply
\be{BO} 
H_{BO} = \av{ H} = \frac 1 {2M} \left( \vec P + e \vec A \right)^2 + \frac 1 {2M} \left(\mu B\right)^2 \av{ \bar z z} + E_0^f,
\ee
\ie a particle with mass $M$ and charge $e=e_1+e_2$ as discussed in the main text. 

For completeness, we also record the exact (unnormalized) LLL wave functions in the case of parabolic confinement potential $V(r) = \half m \omega^2 r^2$:
\begin{align}
  \psi_{L,l} &= Z^L z^l \rme^{-|Z|^2/4\ell^2}\rme^{-\frac{\mu B}{2}\bar Z z }\rme^{-|z|^2/4\ell_{eff}^2},
\end{align}
where the angular momenta $L,l\geq 0$, $\ell = |eB|^{-\half}$ and $\ell_{eff} = (m \sqrt{|e_{eff}B/m|^2+4\omega^2})^{-\half}$. The ground state has $l=0$, yet it is not an eigenstate of the center-of-mass angular momentum, but rather a coherent superposition. A state with a definite center-of-mass angular momentum can be obtained by projecting onto a state with a definite internal angular momentum, say $s$. The projected wave function is
\begin{align}
  \psi_L^{(s)} &= \bar Z^s Z^L\rme^{-|Z|^2/4\ell^2},
\end{align}
where $s$ can now be understood as orbital spin of the center-of-mass state.
\end{appendix}

\end{document}